\documentclass[journal]{IEEEtranTIE}
\usepackage{graphicx}
\usepackage{cite}
\usepackage{picinpar}
\usepackage{amsmath}
\usepackage{url}
\usepackage{flushend}
\usepackage[latin1]{inputenc}
\usepackage{colortbl}
\usepackage{soul}
\usepackage{multirow}
\usepackage{pifont}
\usepackage{color}
\usepackage{alltt}
\usepackage[hidelinks]{hyperref}
\usepackage{enumerate}
\usepackage{siunitx}
\usepackage{breakurl}
\usepackage{epstopdf}
\usepackage{pbox}
\usepackage{algorithm}
\usepackage{algorithmic}

\begin{document}
\title{Physical Unclonable Function-based Key Sharing via Machine Learning for IoT Security}

\author{
	\vskip 1em
	{
	Jiliang Zhang, \emph{IEEE Senior Member},
	Gang Qu, \emph{IEEE Senior Member}

	}

\thanks{Manuscript received 22-Jan-2018; revised 01-Apr-2019 and 23-Jun-2019; accepted 15-Aug-2019. Date of publication 2020; date of current version 2019. This work is supported by the National Natural Science Foundation of China under Grant NO. 61874042 and 61602107, the Key Research and Development Program of Hunan Province under Grant No. 2019GK2082, the Hu-Xiang Youth Talent Program under Grant No. 2018RS3041, and the Fundamental Research Funds for the Central Universities. (\textit{Corresponding author: Jiliang Zhang})
}

\thanks{J. Zhang is with the College of Computer Science and Electronic Engineering, Hunan University, Changsha 410082, China, and also with Cyberspace Security Research Center, Peng Cheng Laboratory, Shenzhen, China (e-mail: zhangjiliang@hnu.edu.cn).}
\thanks{G. Qu is with the Department of Electrical and Computer Engineering,University of Maryland, College Park, USA.}

}

%
%
%
%

\maketitle
	
\begin{abstract}
In many Industry Internet of Things (IIoT) applications, resources like CPU, memory, and battery power are limited and cannot afford the classic cryptographic security solutions. Silicon Physical Unclonable Function (PUF) is a lightweight security primitive that exploits manufacturing variations during the chip fabrication process for key generation and/or device authentication. However, traditional weak PUFs such as Ring Oscillator (RO) PUF generate chip-unique key for each device, which restricts their application in security protocols where the same key is required to be shared in resource-constrained devices. In order to address this issue, we propose a PUF-based key sharing method for the first time. The basic idea is to implement one-to-one input-output mapping with Lookup Table (LUT)-based interstage crossing structures in each level of inverters of RO PUF. Individual customization on configuration bits of interstage crossing structure and different RO selections with challenges bring high flexibility. Therefore, with the flexible configuration of interstage crossing structures and challenges, CRO PUF can generate the same shared key for resource-constrained devices, which enables a new application for lightweight key sharing protocols.
\end{abstract}


\markboth{IEEE TRANSACTIONS ON INDUSTRIAL ELECTRONICS}%
{}

\definecolor{limegreen}{rgb}{0.2, 0.8, 0.2}
\definecolor{forestgreen}{rgb}{0.13, 0.55, 0.13}
\definecolor{greenhtml}{rgb}{0.0, 0.5, 0.0}

\section{Introduction}

\IEEEPARstart{W}{ith} the increasing demands of security, privacy protection, and trustworthy computing, key generation and device authentication become two of the most challenging design concerns, particularly for systems such as smart cards, sensors, smart phones and industrial internet of things where the lack of persistent power limits the duration of countermeasure enforcement. Traditional security mechanisms store secret keys in electrically erasable programmable read-only memory (EEPROM) or battery-backed non-volatile static random access memory (SRAM), and combine cryptographic algorithms to implement information encryption and authentication. In order to secure cryptographic key storage, tamper-resistant devices with a number of countermeasures to defeat various kinds of physical attacks are developed. However, in many IoT applications, resources like CPU, memory, and battery power are limited so that they cannot afford the classic cryptographic security solutions. Silicon physical unclonable function (PUF) emerged as a new hardware primitive provides a unique device-dependent mapping from challenges to responses based on the unclonable properties of the underlying physical device for device authentication and key generation. The key generated by PUF can resist tampering attacks because the underlying nano-scale structural disorder will most likely be damaged during physical tampering \cite{Zhang2015}. Therefore, PUF is a promising security primitive for Internet of Things.

There has been more than a decade of intensive study on PUFs since it was introduced in the research community \cite{Zhang2014A,Nguyen2014}. Among PUFs of different forms, silicon PUFs are of the most interest in terms of fabrication cost and readiness to be integrated into computing and communication devices. Current silicon PUFs can be classed into strong PUFs and weak PUFs. Strong PUFs can provide a large amount of unique challenge-response pairs (CRPs) for device authentication. On the other hand, weak PUFs exhibit only a small number of CRPs and may not applicable to authentication protocols. The responses of weak PUFs can be used as a device-unique key or seed for conventional encryption systems, while maintaining PUF's advantages such as physical unclonability \cite{Zhang2014A}. In such scenario, the responses will not be exposed to any user or attacker. Arbiter PUF\cite{Lim2005Extracting} and Composite PUF \cite{Sahoo2014} are typical strong PUFs. SRAM PUF \cite{Guajardo2007,Holcomb2009,Liu2018} and Glitch PUF \cite{zhang2013,zhang2016} are typical weak PUFs. Ring oscillator (RO) PUF \cite{Suh2007Physical} only produces a limited number of CRPs which are not large enough for authentication and hence is more suitable for key generation. In addition, an RO PUF does not require high symmetry and thereby is more easily to be implemented on FPGAs than other PUFs such as Arbiter PUF. In past decades, PUFs have attracted much attention in academia and industry and have enabled a variety of security protocols such as authentication \cite{Suh2007Physical,Majzoobi2012,Zhang2018c,Zheng2019,Verayo2017} and encryption/decryption \cite{Suh2007Physical,Zhang2014d,zhang2018}. However, current weak PUFs exhibit a shortcoming when they are used in some security protocols. They generate the chip-unique key for each device and cannot be cloned in another device due to process variation, while some security protocols such as multi-party communication require many parties to share the same key. Therefore, current weak PUFs are inapplicable to such application scenarios.
In order to address this issue, we propose the first PUF-based key-sharing method where the PUF has flexible configuration of challenges and inter-stage crossover structures so it can generate the shared key. We also use a standard key-sharing protocol to show how the shared key can be communicated among adjacent resource-constrained nodes. Our key contribution is on the cost-efficient generation of shared-key which will be elaborated in this paper. Once the shared keys are generated, standard key-sharing protocols can be used.


%

The rest of this paper is organized as follows. We first explain the rationale of using configurable PUFs for shared key generation and review the existing work on configurable PUFs in Section II. The proposed PUF-based key-sharing is elaborated in Section III. Potential security threats and countermeasures are analyzed in Section IV. The detailed experimental results and analysis are reported in Section V. Finally, we conclude in Section VI.

\section{Related Work}

\subsection{PUFs for Shared-key Generation}
In IoT, sensitive information needs to be transmitted to participants over a potentially insecure communication. Hence, security features such as authentication and encrypted data transfer are required. However, it is difficult to secure IoT with security features used in traditional Internet. In order to fit such application scenario, the deployed security features must be extremely lightweight \cite{Qu2014}. Instead of relying on heavyweight public-key primitives or secure storage for secret symmetric keys, PUF is a lightweight hardware primitive that can be directly integrated in cryptographic protocols. So far, all existing PUF-enable encryption/decryption protocols follow the same paradigm: PUFs generate the chip-unique key for each resource-constrained device and cannot be shared securely in another resource-constrained device.

In this paper we consider application scenarios where PUF-enable encryption/decryption schemes fail to work: multi-party communication needs to share the same key. To our knowledge, PUFs for shared-key generation have not been reported in current references. We construct the first and efficient PUF-based security protocol for this setting. Therefore, this is the first work that PUFs can generate the same lightweight shared-key in physically for resource-constrained devices.

In multi-part communication, all the nodes need to have the same encryption/decryption keys. If weak PUF is used to generate such key, we cannot guarantee that they will create the same key. Indeed, that would be against the uniqueness principle of PUF [3]. On the other hand, strong PUFs can provide a large number of CRPs. However, the challenge response pair works similar to a one-way hash. It is extremely difficult, if not impossible to find a challenge which can generate a required response (i.e. the shared key). If a trusted third party (TTP) knows all the CRPs for all the strong PUFs, then it is possible to find a corresponding challenge for each PUF so they can have the same response. However, this approach is impractical given the large size of the CRPs and the number of nodes that want to share a key. In short, weak PUF does not have sufficient key space and strong PUF based solution is impractical. We propose a novel practical approach based on the recently proposed configurable PUF \cite{Pang2017} to solve this problem. The configurable PUF can be controlled conveniently to generate the same key on multiple nodes and is capable of providing the sufficient number of shared keys for IoT applications. Next, we review the existing work on configurable PUFs.
\subsection{Configurable PUFs}


\begin{figure}[!t]
\centering\includegraphics[width=\linewidth]{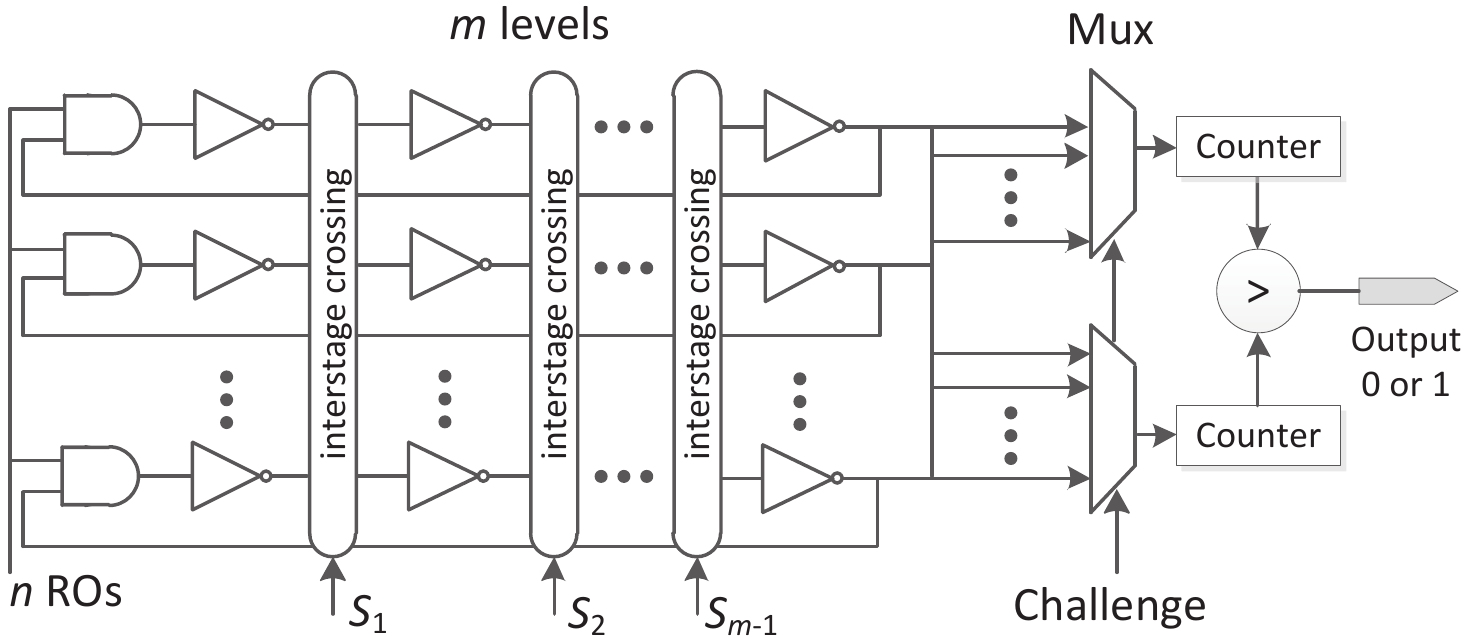}
\caption{Crossover RO PUF structure \cite{Pang2017}}
\label{fig:3}
\end{figure}

%
Configurable RO PUF is introduced by Maiti and Schaumont \cite{Maiti2011Improved} to improve RO PUF reliability. The key idea is that a $2\times1$ multiplexer is used to select one out of two inverters at each stage of the RO. This technique uses the configurations with the largest delay difference to improve the PUF reliability. Another highly flexible configurable RO PUF was proposed in \cite{Gao2014A}. The key idea is that a $2\times1$ multiplexer is used to select or bypass the inverter to improve the reliability of RO PUFs. The configurations for RO pairs are to choose the largest delay difference to generate reliable PUF output. For these reconfigurable/configurable PUFs, the utilization ratio of multiplexers added in ROs is low, and the inverters are not fully used in some configurations.

Recently, Pang et al. \cite{Pang2017} proposed a crossover RO PUF architecture which shows the higher flexibility of selecting inverters in ROs. The crossover RO PUF has \emph{n} ROs and \emph{m} levels of inverters. Each RO consists of \emph{m} inverters with a particular frequency. The $m$ should be larger than 2, otherwise, the RO would oscillate too fast to be precisely counted by the counter. For \emph{m} levels of inverters, the outputs of previous inverter level are fed as the inputs to the next inverter level after interstage crossing. The interstage crossing cell determines the routing path of step signals input without any additional logical operation. There are \emph{m}-1 interstage crossing cells to change the configuration of the delay loop with selection inputs. As shown in Fig. \ref{fig:3}, configuration selection $S=(S_1, S_2, ... , S_i, ... , S_{m-1})$, where $S_i$ has $\left \lceil \\log_2 n\right \rceil$ bits and determines the connection order of inverter to the next inverter level in \emph{i}-th stage; The configuration selection $S$ and challenge are combined together as the whole challenge to be input into the CRO PUF for generating the response. $S_{m-1}$ is dedicated to ensure closed loops. The number of possible different configurations of the delay loops is $(A_n^n)^{m-2}$. The level \emph{m} must be an odd number and $m > 2$ in order to make the delay loop form the oscillation, and it can determine the frequency level of the RO. The \emph{m} is not directly related to \emph{n}. The \emph{n} determines the number of possible challenges, while the \emph{m} determines the frequency level of the ROs. The larger \emph{m} which means more inverters in RO exhibits lower frequencies. In practical applications, the frequencies of ROs should not be too high and too low. If the frequency is too high, high-precision counter is required; if the frequency is too low, the time to generate response would be long, and hardware and power overhead would be increased. Usually, we can set $m$ to 5 or 7 which is the empirical value that meets above requirements.

\section{PUF-based Key-sharing}
\subsection{Principle of Shared-key Generation}
The shared key is required in multi-party communication between different devices. Traditional PUFs generate chip-unique key for every device, while CRO PUF is able to generate the same shared key for all devices. Therefore, configurable PUFs \cite{Maiti2011Improved,Gao2014A,Pang2017} are suitable for one-to-many authentication. PUF-based shared-key generation can be implemented by all configurable PUFs. This paper takes CRO PUF \cite{Pang2017} as an example for shared-key generation. CRO PUF is based on the inter-stage crossover structure which can be configured with the SRAM value. Different devices can produce the same response as the shared key with the appropriate configurations and challenges. For a CRO PUF with \emph{n} rows and \emph{m} columns, there are (\emph{m}-1) $\left \lceil \log_2 n \right \rceil$-bit selection signals which have (A$_n^n$)$^{m-2}$ combinations. The challenge \emph{C} can have up to A$_n^2$  different selections with the multiplexers to select any two ROs for frequency comparison. With the increasing of \emph{n} and \emph{m}, the number of the selection signals and the challenges increases exponentially. In addition, the number of configurations of inter-stage crossover structures provides high flexibility for one-to-many authentication.
The delay model of a CRO PUF with \emph{n} rows with \emph{k} inverters is shown as follows.\\

$ Delay_{RO}=\begin{matrix}
     d_{11} & d_{12} & \ldots & d_{1j} & \ldots & d_{1k}\\
     d_{21} & d_{22} & \ldots & d_{2j} & \ldots & d_{2k}\\
     \vdots & \vdots &        & \vdots &        & \vdots\\
     d_{i1} & d_{i2} & \ldots & d_{ij} & \ldots & d_{ik}\\
     \vdots & \vdots &        & \vdots &        & \vdots\\
     d_{n1} & d_{n2} & \ldots & d_{nj} & \ldots & d_{nk}\\
\end{matrix}$
\\

The delay vector of each line $D_{RO}=\{D_1,D_2,\ldots,D_i,\ldots,D_n\}$, where $D_i=\sum_{j=1}^k d_{ij}$.
The selection signal $ S=(S_1,S_2,\ldots,S_j,\ldots,S_{k-1})$, where $ S_j $ controls the connection path between the \emph{j}-th and (\emph{j}+1)-th column inverters, i.e,
             \begin{center}
              $ \{d_{1j}',d_{2j}',\ldots,d_{nj}'\} =f(d_{1j},d_{2j},\ldots,d_{nj})$ \\
             \end{center}

The challenge C$_{RO}$ is used to choose different rows of ROs for the frequency comparison, i.e,
\begin{center}
$\{D_1',D_2'\}=g(D_1,D_2,\ldots,D_n)$
\end{center}

The selection signal \emph{S} adjusts the delay of each column with the function \emph{f}. Challenge uses the function \emph{g} to select different rows of ROs for comparison to generate the response. The delay is different between any two CRO PUFs, but we can get the same response by using function \emph{f} and \emph{g} with different C$_{RO}$ and \emph{S}. Function \emph{f} and \emph{g} are independent. The function \emph{f} is to rearrange the column vectors, and the function \emph{g} is to select the column vector elements. In the one-to-many authentication, \emph{f} and \emph{g} are used for the configuration to get the same response from any two different CRO PUFs. Based on this, we can design a shared pairing key generation scheme. In what follows, we give an example to explain the idea.

Taking two CRO PUFs as an example, each CRO PUF contains four 4-layer inverters. The corresponding delay models are represented by matrices \emph{A} and \emph{B}, respectively.
\begin{center}
$ A=\begin{matrix}
     3&  6 & 8 & 5 \\
     9 & 7 & 4 & 5 \\
     5 & 4 & 6 & 5 \\
     2 & 5 & 6 & 3 \\
\end{matrix}$
   \ \ \  $ B=\begin{matrix}
     2&  4 & 6 & 5 \\
     5& 1 & 3 & 2 \\
     8& 6 & 5 & 7 \\
     3 & 6& 4& 5 \\
   \end{matrix}$
\end{center}

Consider the following challenges:
\begin{center}
$C_A=\{\{1,2\},\{2,3\},\{3,4\}\}$\\
$C_B=\{\{1,3\},\{3,4\},\{4,2\}\}$\\
\end{center}

In this case, the responses of both CRO PUFs are \{0,1,1\}. Similarly, assuming the challenges are
\begin{center}
$C_A=\{\{4,2\},\{2,3\},\{3,1\}\}$\\
$C_B=\{\{1,2\},\{2,4\},\{4,3\}\}$\\
\end{center}

Keep the selection signals \emph{S} in \emph{A} unchanged and adjust $S_2$ and $S_4$ in \emph{B}, the path delay models become
\begin{center}
$ A=\begin{matrix}
     3&  6 & 8 & 5 \\
     9 & 7 & 4 & 5 \\
     5 & 4 & 6 & 5 \\
     2 & 5 & 6 & 3 \\
\end{matrix}$
   \ \ \  $ B=\begin{matrix}
     2&  4 & 6 & 5 \\
     5& 6 & 3 & 7\\
     8& 1 & 5 & 5 \\
     3 & 6& 4& 2 \\
   \end{matrix}$
\end{center}

The delays of \emph{A} and \emph{B} become
\begin{center}
$D_A=\{22,25,20,16\}$\\
$D_B=\{17,21,19,15\}$\\
\end{center}

In this case, the responses of both CRO PUFs are \{0,1,0\}.


It is important to mention here that the delay of the crossover structure may also impact the total delay of the PUF. Early works \cite{Qu2009,Majzoobi2010FPGA} have pointed out that existing FPGA design tools can minimize the delay-skew between a pair of routes but cannot guarantee exact symmetric structure for the identical delay. For example, the multiplexers and inverters in a configurable RO PUF (Fig. \ref{fig:3}) will be connected to the switch matrix which uses routes with different lengths depending on the individual placements. Therefore, if the manufacturing process variation on the logic (i.e. multiplexers and inverters) is insufficient to offset the delay difference caused by the routes, the interstage crossing structure could make the PUF circuit biased. Fortunately, \cite{Pang2017} have demonstrated that crossover RO PUF can set all connections of inverters by configuring SRAM without affecting the routing in the switch matrices. Indeed, a high uniqueness (49\%) was reported in \cite{Pang2017} which means that delay of the crossover circuits has little impact. However, if one can build an accurate model for the crossover structure and integrate it into the delay model for the PUF, the uniqueness and robustness could be improved.

\begin{figure}[!t]
\centering\includegraphics[width=\linewidth]{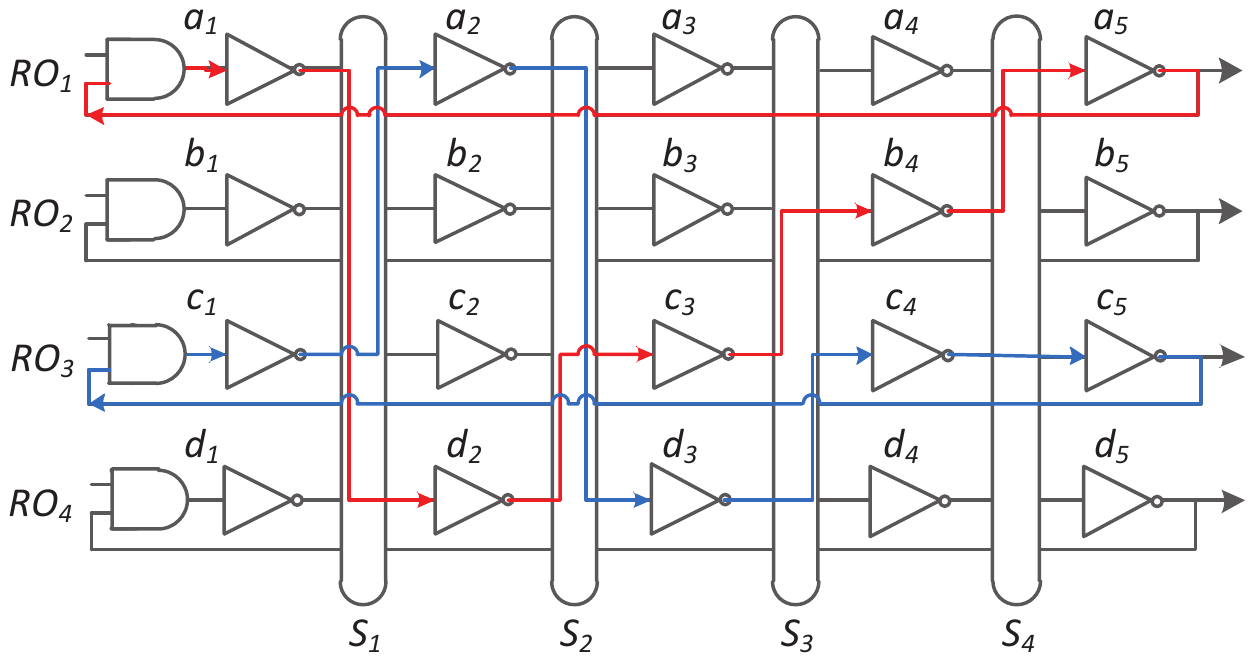}
\caption{An example of two delay paths}
\label{fig:12}
\end{figure}

\subsection{Modeling of Delay Matrix}
We used machine learning algorithms to model the CRO PUF and get the required delay matrix. The real delay matrix of the PUF that is implemented on hardware is called the original delay matrix, and the delay matrix we obtained through modeling is called the predicted delay matrix.

In the real scenario, it is difficult to get the original delay matrix, but all delay paths and their corresponding numbers in the counters can be measured on the CRO PUF. Therefore, the predicted delay matrix can be generated with the following two steps:
\begin{enumerate}
  \item Enumerate all delay paths and get the numbers in the counters on the original delay matrix.
  \item Build a model to obtain the predicted delay matrix.
\end{enumerate}

Generally, the PUF responses are generated with the challenges that are used to select any two delay paths to compare. For example, as shown in Fig. \ref{fig:12}, the delay difference between the two paths (marked as red and blue) is $(a1 + d2 + c3 + b4 + a5) - (c1 + a2 + d3 + c4 + c5)$. However, in the actual model, the delay parameters are difficult to be predicted using original CRPs. Therefore, during the delay predicting phase, we use the numbers in the counters to compute the exact values of the delay of each configuration, named as $Counts$. For example, the configuration of the red path ``$(a1 + d2 + c3 + b4 + a5)$'' in Fig. \ref{fig:12} and the parameters of the delay matrix $W$ can be shown as follows.

\begin{center}
$ C=\begin{matrix}
     1&  0 & 0 & 0 & 1\\
     0 & 0 & 0 & 1 & 0\\
     0 & 0 & 1 & 0 & 0\\
     0 & 1 & 0 & 0 & 0\\
\end{matrix}$
   \ \ \  $ W=\begin{matrix}
     a_1 &  a_2 & a_3 & a_4 & a_5 \\
     b_1 &  b_2 & b_3 & b_4 & b_5 \\
     c_1 &  c_2 & c_3 & c_4 & c_5 \\
     d_1 &  d_2 & d_3 & d_4 & d_5 \\
   \end{matrix}$
\end{center}

$C \cdot W$ is the dot product of the matrix $C$ and $W$, so the number $Counts$ in the counter given by:

\begin{equation}
Counts=\frac{1}{C \cdot W}
\end{equation}

All delay paths and their corresponding numbers in the counters can be enumerated for a CRO PUF. Therefore, machine learning algorithms can be used to fit the parameter \emph{W} to get the  predicted delay matrix of CRO PUF. In this case, the input-output behavior of CRO PUF is completely consistent with the predicted model.

Note that there is a counter access interface implemented by fuses in the PUF so that the designer can obtain the numbers in the counters to model the PUF, and then burn the fuses to destroy the access interface before distributing the chips for usage. In this way, designers can model the PUF while attackers are prohibited.

\subsection{Shared-key Generation}

\begin{algorithm}[!t]
\renewcommand{\algorithmicrequire}{\textbf{Input:}}
\renewcommand{\algorithmicensure}{\textbf{Output:}}
\caption{\textbf{Get the threshold}}
\label{alg1}
\begin{algorithmic}[1]
\REQUIRE Delay matrix $A$, Empty set $B$
\ENSURE Threshold $T$
\STATE \textbf{Initialize:} $D$: the set of all possible path delay in $A$
\FOR {$d_1 \in D$, $d_2 \in D$ \textbf{and} Path($d_1$) $\ne$ Path($d_2$)}
\STATE $B$ $\gets$ {($|d_1 - d_2|$, Config($d_1, d_2$))} $\cup$ $B$
\STATE $//$Config($d_1$, $d_2$): the configuration challenge of $d_1$ and $d_2$
\ENDFOR
\STATE Sorted($B$) $//$sorted by descending order according to the absolute value of delay difference
\STATE $N$ = Size($B$)
\FOR {$i=1, 2, \ldots, N$ \textbf{and} ($X_i$, $Y_i$) $\in$ $B$}
\STATE $//$$X_i$: the absolute value of the delay difference
\STATE $//$$Y_i$: the configuration challenge
\IF {$Y_i$ generates an unstable response at different temperatures}
\STATE $T$ = $X_i'$ $//$$X_i'$ is a reasonable value greater than $X_i$
\STATE \textbf{break}
\ENDIF
\ENDFOR
\RETURN $T$
\end{algorithmic}
\end{algorithm}


\subsubsection{Reliable response Generation for Shared-key}
Shared-key generation requires CRO PUF generating stable responses. As discussed above, we can get a high accuracy predicted delay matrix already. Therefore, the delay difference between any two paths can be obtained easily. On this basis, we sort the absolute values of the delay differences between all the paths by descending order, and take into account the influence of different temperatures to determine a threshold $T$. When the absolute value of the delay difference between the two paths is greater than the threshold, the response of the two paths can be considered stable even under different temperatures. The selection of threshold shows as the \texttt{Algorithm 1}. In this algorithm, we store the absolute value of delay difference and configuration challenge between all paths in the set \emph{B}. Then we sort the elements in \emph{B} by descending order according to the absolute value of delay difference. Finally, we enumerate the elements in \emph{B} to determine whether the configuration challenge can generate stable response at different temperatures. If a configuration challenge does not produce a stable response, we will use the absolute value of the delay difference between the two paths as the threshold $T$, and the threshold would be increased to ensure that a stable response is generated. The detailed explanation for \texttt{Algorithm 1} is as follows.

\begin{itemize}
  \item Store all paths of delay matrix $A$ into the set $D$ (line 1);
  \item Enumerate all combinations of two different paths in the set $D$ (line 2-5);
  \item Store the absolute value of the delay difference, configuration and challenge of two paths into the set $B$ as an element (line 3);
  \item Sort the elements in the set $B$ by descending order according to the absolute value of delay difference (line 6);
  \item Enumerate the elements in the set $B$ and determine whether the response generated by the corresponding configuration and challenge keeps stable at different temperatures (line 8-15);
  \item If stable, continue to enumerate the elements in the set $B$. Otherwise, use the absolute value of the current delay difference as the appropriate threshold $T$ and end the loop. Note that, in practice, the threshold will be increased slightly to ensure response 100\% reliable (line 11-14).
\end{itemize}

\begin{algorithm}[!t]
\renewcommand{\algorithmicrequire}{\textbf{Input:}}
\renewcommand{\algorithmicensure}{\textbf{Output:}}
\caption{\textbf{Challenge generation}}
\label{alg1}
\begin{algorithmic}[1]
\REQUIRE Delay matrix $A$, Threshold $T$, Shared-key $K$
\ENSURE Challenge $C$
\STATE \textbf{Initialize:} $D$: the set of all possible path delay in $A$
\STATE $N$ = Size($K$)
\FOR {$i=1, 2, \ldots, N$}
\WHILE {$d_1$ $\gets$ random($D$), $d_2$ $\gets$ random($D$) \textbf{and} Path($d_1$) $\ne$ Path($d_2$)}
\STATE $//$random($D$): select a random path delay in $D$
\IF{$K_i$ == 1 \textbf{and} $d_1 - d_2$ $>$ $T$}
\STATE $//$$K_i$ is the $i$-th bit of $K$
\STATE $C_i$ = Config($d_1, d_2$) $//$$C_i$ can generate $K_i$
\STATE \textbf{break}
\ELSIF{$K_i$ == 0 \textbf{and} $d_1 - d_2$ $<$ $-T$}
\STATE $C_i$ = Config($d_1, d_2$)
\STATE \textbf{break}
\ENDIF
\ENDWHILE
\ENDFOR
\STATE $C$ = \{$C_1$, $C_2$, \ldots, $C_N$\}
\RETURN $C$
\end{algorithmic}
\end{algorithm}


Unlike configurable RO PUF designs which are only to make the response stable, our design aims to extract the stable response as the shared key on the IoT devices. To achieve this purpose, we propose the challenge configuration generation for the RO, which is illustrated in \texttt{Algorithm 2} and explained in the following. Hot carrier injection, electro-migration, negative bias temperature instability and temperature dependent dielectric breakdown cause the device aging which is another cause of the reliability problem for PUFs. In recent years, there are several papers focusing on this topic \cite{Maiti2011} and also several aging-resistant techniques are developed \cite{Rahman2015}. We did not propose an aging-resistant technique for the RO PUF in this paper, but it is worth developing new aging-resistant techniques to further improve the reliability of PUFs in future work.

\subsubsection{Challenge Generation for Shared-key}
In the key-sharing protocol which we will introduce in the next Section, a trusted third party (TTP) possesses the delay matrix of CRO PUF and the threshold for generating a stable response. The TTP needs to generate the challenges of CRO PUFs corresponding to the key that needs to be shared. In this Section, we propose a heuristic challenge generation algorithm shown in \texttt{Algorithm 2}. In the \texttt{Algorithm 2}, for each bit of the key, the TTP will randomly select two paths and determines whether their delay differences are greater than the threshold. If Yes, TTP will get configuration challenge for these two paths, otherwise TTP would reselect another two paths randomly to compute the challenge. The detailed explanation for \texttt{Algorithm 2} is as follows.

\begin{itemize}
  \item Store all paths of delay matrix $A$ into the set $D$ (line 1);
  \item Enumerate all the bits of the shared-key $K$ (line 3-15);
  \item Randomly select two different paths from the set $D$ (line 4-14);
  \item For each bit of the shared key $K_i$, if $K_i$ equals 1 and the delay difference is greater than $T$, it means that we found a configuration challenge $C_i$ that can generate a stable response 1 (line 6-9); If $K_i$ equals 0 and the delay difference is less than $-T$, it means that we found a configuration challenge $C_i$ that can generate a stable response 0 (code line 10-12); Otherwise, return to step 2.
  \item The configuration challenge $C$ is composed of all $C_i$ (line 16).
\end{itemize}

\begin{figure}[!t]
\centering\includegraphics[width=\linewidth]{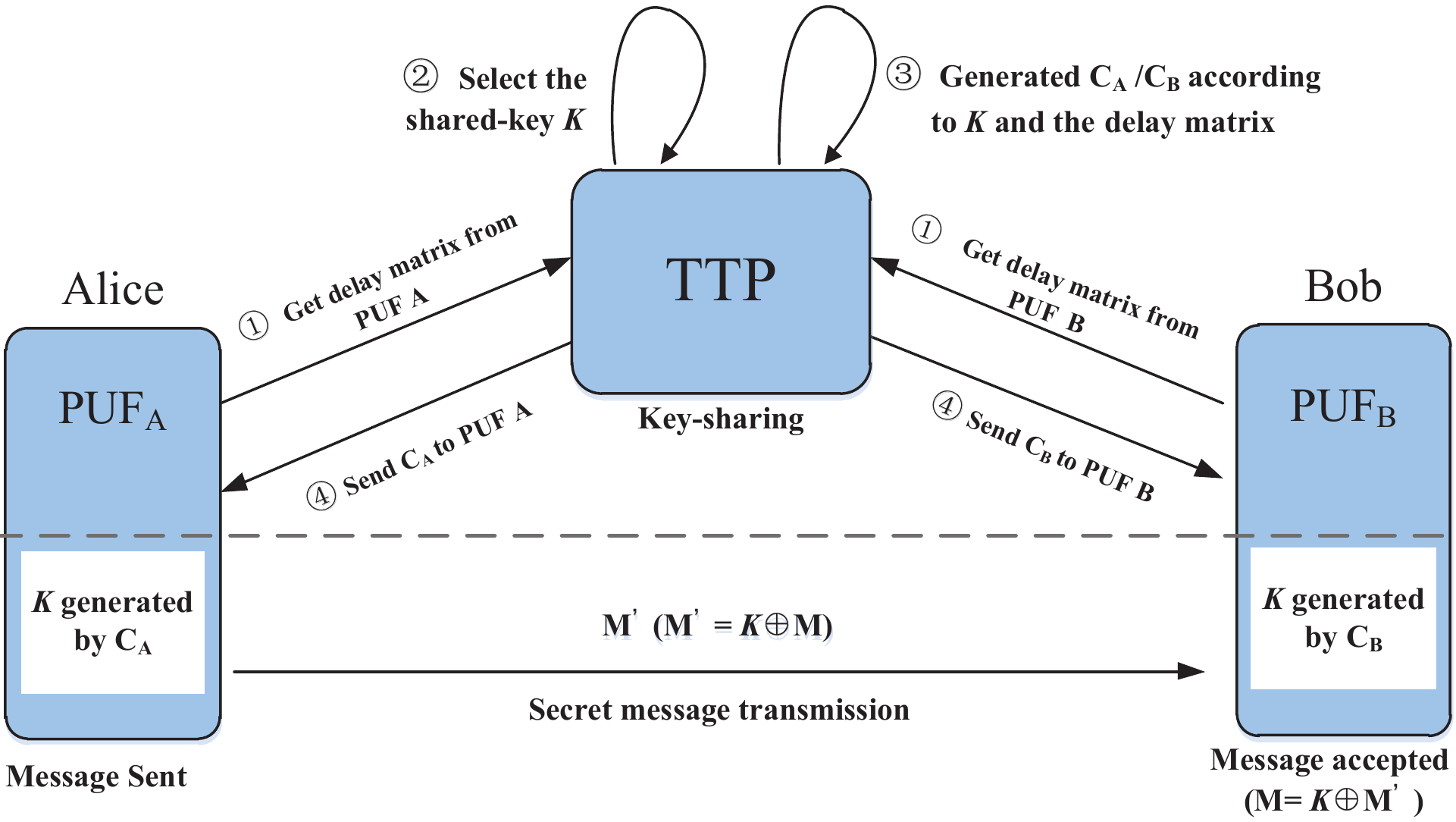}
\caption{PUF-based key-sharing and secret information transmission protocol}
\label{fig:11}
\end{figure}

\subsection{Key-sharing Protocol}

As discussed above, CRO PUF can generate the shared key with the flexible configuration of \emph{S} and \emph{C}. Therefore, it can be used as authentication of multi-party communication for adjacent resource-constrained nodes. As shown in Fig. \ref{fig:11}, we propose a CRO PUF-based key-sharing and secret information transmission protocol.

Assuming that $PUF_A$ and $PUF_B$ require sharing the key. First, we send the delay matrix of $PUF_A$ and $PUF_B$ to the TTP, and in this case, the TTP carries the delay matrix of $PUF_A$ and $PUF_B$. Second, the TTP selects a key \emph{K} that needs to be shared between $PUF_A$ and $PUF_B$. Third, TTP generates the challenge $C_A$ and $C_B$ according to the delay matrix of $PUF_A$, $PUF_B$ and the shared-key \emph{K}. Then TTP sends $C_A$ and $C_B$ to $PUF_A$ and $PUF_B$, respectively. Finally, $PUF_A$ and $PUF_B$ are able to generate the shared-key \emph{K} with $C_A$ and $C_B$, respectively. In the whole process, there is no secret key transmission. Besides, configuration information $S$ is not the secret information and can be stored in SRAM. Therefore, it is with high security and low cost to realize the key-sharing among multi-parties.

After $PUF_A$ and $PUF_B$ have obtained the shared-key, they can transfer the secret message to each other. For example, we suppose that Alice needs to send message \emph{M}(10100101) to Bob. At this time, Alice encrypts the message \emph{M} by XOR with the secret key \emph{K}(01101001) to generate the encrypted message $M'$(11001100 = $K \oplus M$), and sends $M'$ to Bob. After receiving the $M'$, Bob gets the message \emph{M} (10100101 = $M' \oplus K$) by XOR with the secret key \emph{K}.

\begin{figure}[!t]
\centering\includegraphics[width=\linewidth]{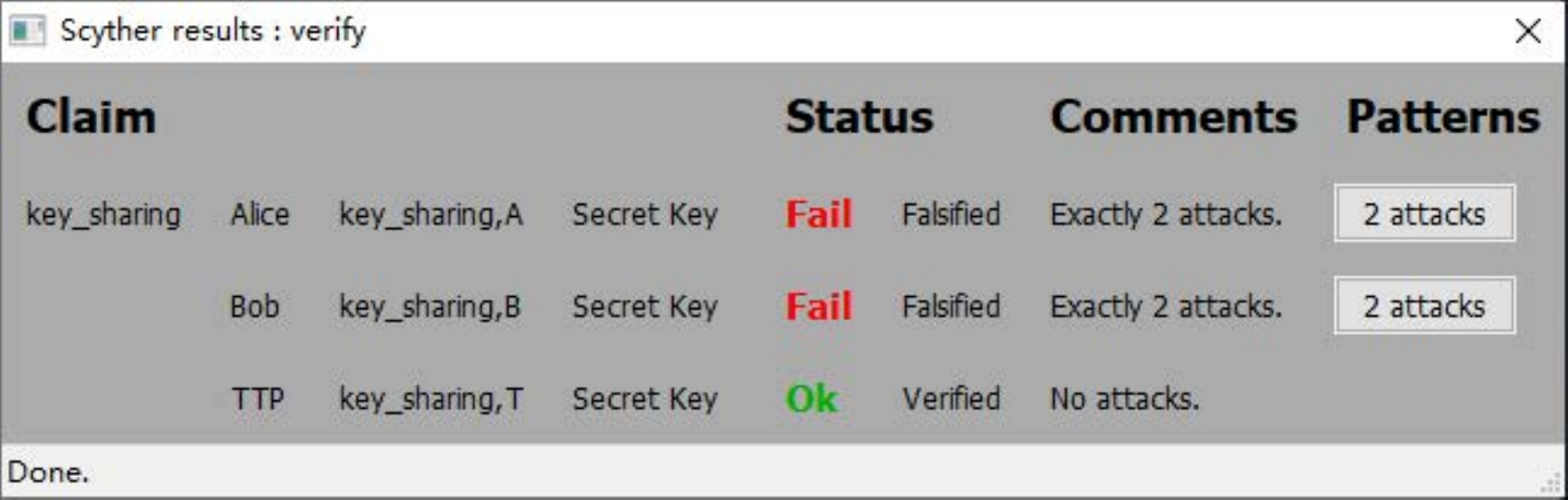}
\caption{The verification results of key-sharing protocol on Scyther.}\label{fig:17}
\end{figure}

As discussed above, a CRO PUF-based key-sharing method and the corresponding secure information transmission protocol are proposed. It is noted that PUF-based key-sharing is not specially designed for RO PUF but also can be applied to other PUF structures, provided that the delay information are obtained by testing and appropriate challenges are selected.

\begin{figure}[!t]
\centering\includegraphics[width=\linewidth]{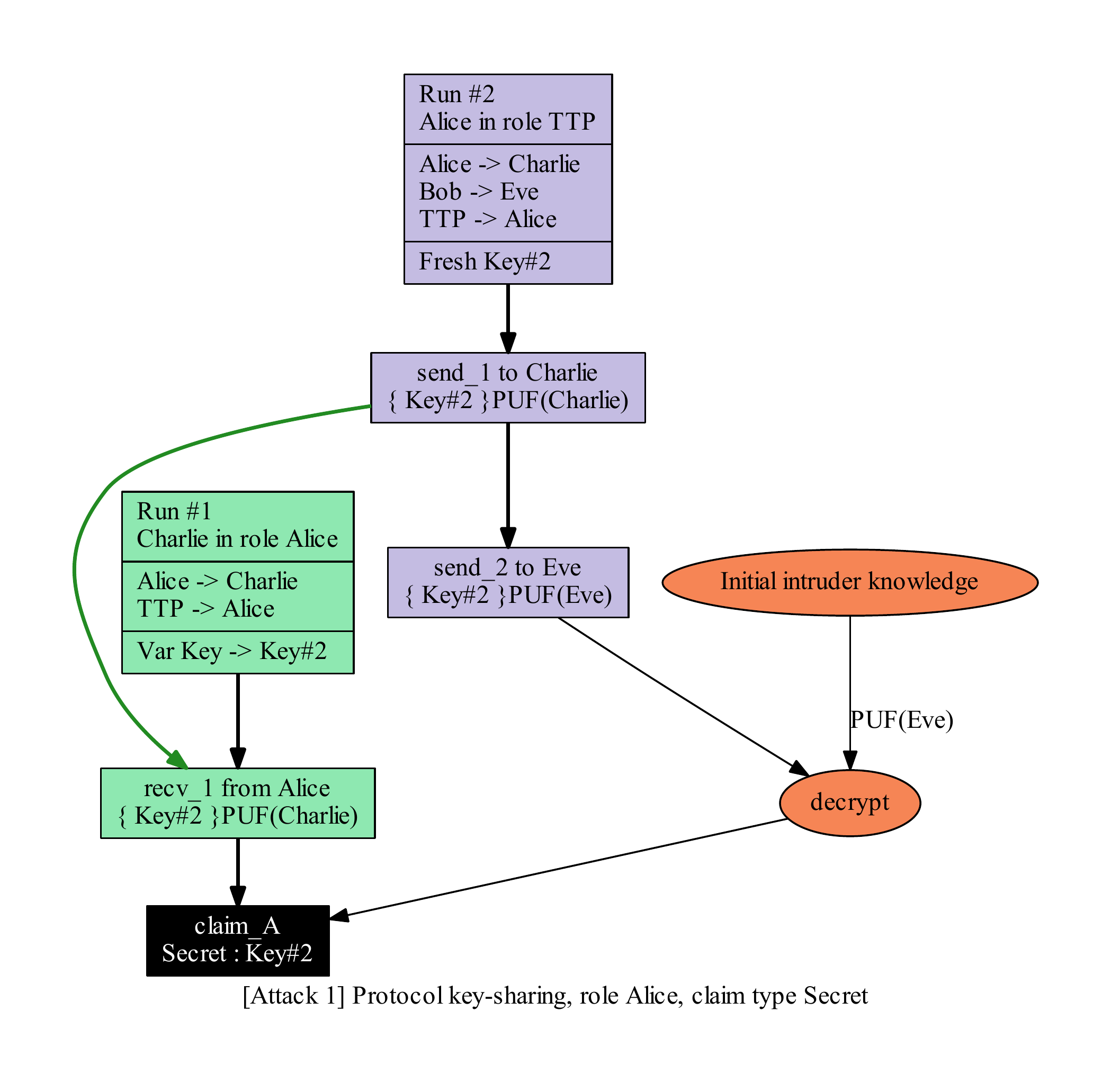}
\caption{Flow of the attack method \#1}\label{fig:18}
\end{figure}

\begin{figure}[!t]
\centering\includegraphics[width=0.8\linewidth]{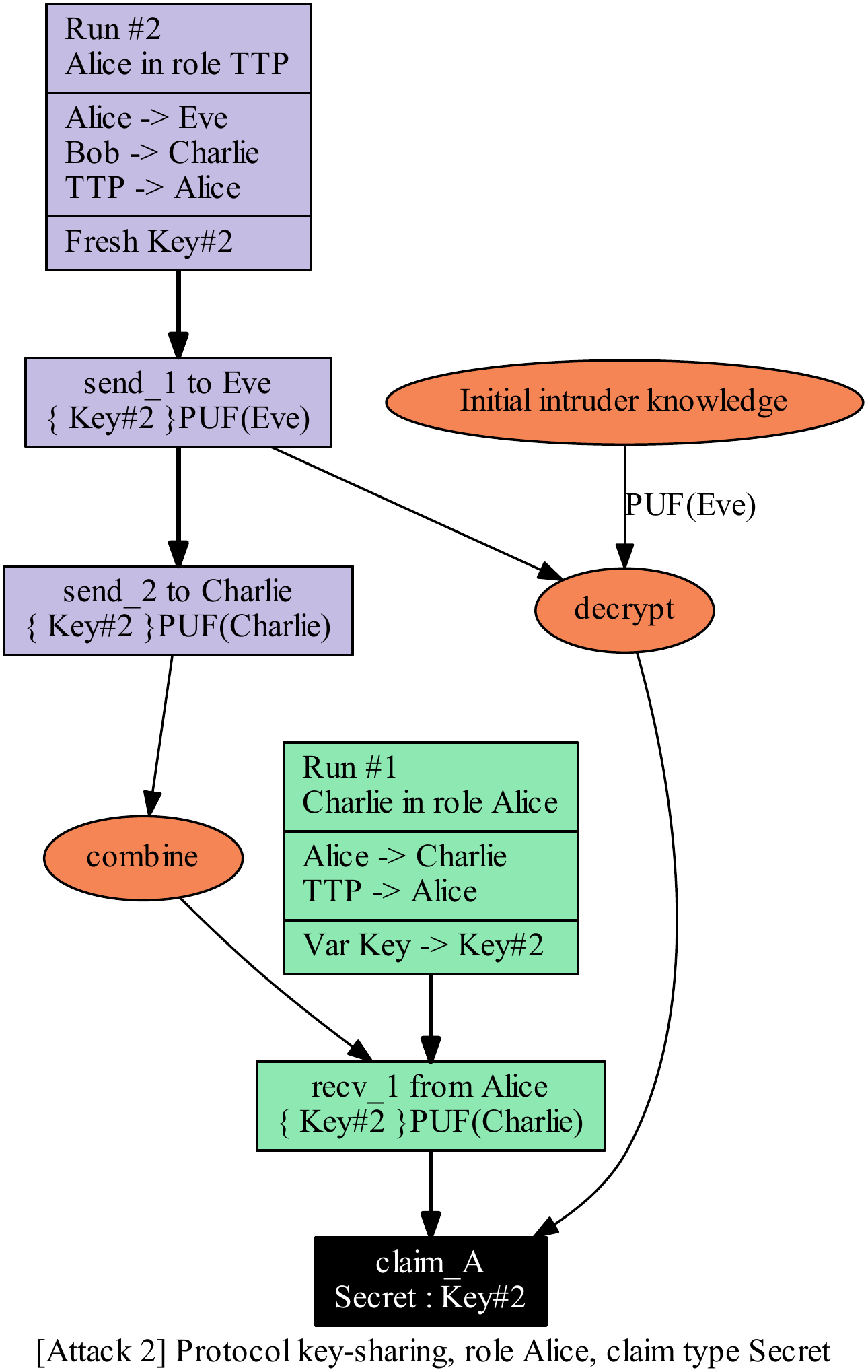}
\caption{Flow of the attack method \#2}\label{fig:19}
\end{figure}

\section{Security Analysis}
The most important feature for physical unclonable function is ``unclonable'' obviously. However, this feature is threatened with the attack techniques reported recently. Machine learning (ML)-based modeling attacks and side-channel attacks are two kinds of main threatens for RO PUF.

\subsection{Modeling attacks}
ML-based modeling attacks are the most efficient attack for strong PUFs which have a publicly accessible CRP interface so that attackers can collect a large number of CRPs to model the PUF with the mathematical way \cite{Zhang2014A,Sahoo2015}. For example, our recent experimental results show that machine learning can foresee arbiter PUF responses to given 1000 CRPs with prediction rates up to 99\%. However, crossover RO PUF is used as the weak PUF for key generation. The corresponding responses are used as keys. In such application, there is no access interface to read the key generated inside the chip so that the key will not be exposed to attackers (CRP access interface is implemented by fuses which will be destroyed after designers obtain the CRPs \cite{Zhou2017}). Therefore, it is difficult to conduct ML-based modeling attacks on the CRO PUF.

\subsection{Side-channel attacks}
Side-channel attacks statistically analyze the time, power consumption or electromagnetic emanation of the cryptographic devices to gain knowledge about integrated secrets. Most recently, Merli et al. carried out side-channel attacks (EM analyses) on an RO PUF FPGA implementation leading to the extraction of a full PUF model and thereby breaking the PUFs security \cite{Merli2013Localized}. The authors also point that their proposed attack can be successful because they exploit that each RO has a fixed location and a specific measurement path through a multiplexer to a counter. In this paper, we can dynamically change the inverters of ROs with different configuration data to generate the updated key, which makes each RO have no fixed physical location and therefore our proposed crossover RO potentially provides a new solution to resist side-channel attacks. Moreover, the security can be enhanced by increasing the number of inverters in ROs and levels of ROs.

\subsection{The security analysis of key-sharing protocol}
We use Scyther protocol analysis tools to analyze the security of the PUF-based key-sharing protocol which is shown in Fig. \ref{fig:11}. The experiment environment is set up with Scyther 1.1.3, graphviz 2.38, Python 2.7 and wxPython 2.8. The verification results are shown in Fig. \ref{fig:17}.

\begin{figure}[!t]
\centering\includegraphics[width=\linewidth]{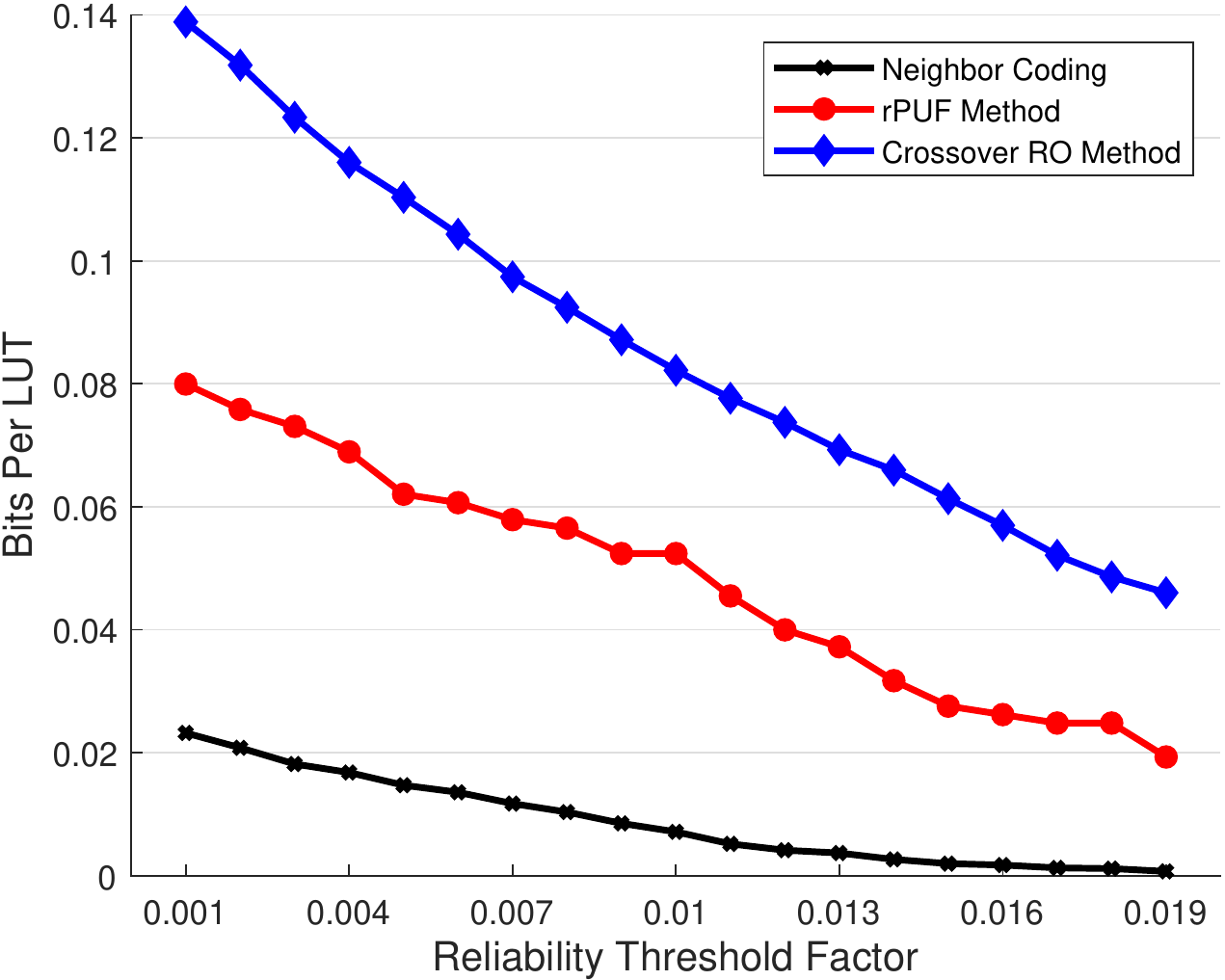}
\caption{Comparison of hardware efficiency}
\label{fig:6}
\end{figure}

As shown in Fig. \ref{fig:18} and Fig. \ref{fig:19}, Scyther gives two possible attack methods to break our protocol to get the key. In the attack method \#1/\#2, it is assumed that the attacker knows Bob/Alice's PUF model. In this case, the attacker intercepts the encrypted information sent to Bob/Alice and use the PUF model for decryption to derive the key. Both the two attack methods assume that the attacker knows the original PUF model or pre-trained PUF model. However, in the key-sharing protocol, the original PUF model can not be cloned and the pre-trained PUF model is stored on the TTP. Therefore, it is difficult to break our proposed key-sharing protocol.

\section{Experimental results}

We implement the traditional neighbor coding RO PUF, rPUF \cite{Gao2014A}, and crossover RO PUF on ZedBoard xc7z020clg484-1 FPGAs to compare the hardware efficiency.

\begin{table} [!t]
\caption{Hardware overhead}\label{tab3}
\begin{center}
\begin{tabular}{|c|c|}
\hline
RO PUFs & No. of LUTs \\
\hline

Neighbor coding RO PUF with 512 ROs & 8704 \\
\hline
$14 \times 5$ rPUF & 322 \\
\hline
$4 \times 5$ Crossover RO PUF & 62 \\
\hline
\end{tabular}
\end{center}
\end{table}

\subsection{Hardware Efficiency}
 As shown in Table \ref{tab3}, a decouple neighbor coding RO PUF uses 512 ROs to form 256 RO pairs to generate 256-bit response, which consumes $8704$ LUTs. A rPUF with 14 5-level ROs consumes $322$ LUTs. A crossover RO PUF with 4 5-level ROs consumes $62$ LUTs. In order to evaluate the hardware efficiency of our proposed CRO PUF, we use the bits per LUT to evaluate the hardware efficiency of each RO PUF. We define the reliability threshold factor \cite{Zhang2014d} which is denoted as $\sigma$. Consider a pair of ring oscillator, $RO_A$ and $RO_B$, and assume their frequencies are $F_A$ and $F_B$, respectively. Then $\mid F_A-F_B \mid\geq F_{TH}=F_{ref}\sigma$, where $F_{ref}$ is the frequency of reference RO, and $\sigma$ is the reliability threshold factor \cite{Zhang2014d}. Fig. \ref{fig:6} denotes the comparison of hardware efficiency of three methods. The hardware efficiency is denoted by the number of bits generated by per LUT. As shown in Fig. \ref{fig:6}, more hardware resource would be used when the criterion on reliability threshold factor for acquiring reliable ones is tightened, and crossover RO PUF is obviously more efficient than the other two methods. For example, when $\sigma$ is set to 0.01, compared with decouple neighbor coding and rPUF, CRO PUF can get 11.54 times and 1.57 times hardware reduction, respectively, when generating the same number of reliable PUF response bits.

\begin{table} [!t]
\newcommand{\tabincell}[2]{\begin{tabular}{@{}#1@{}}#2\end{tabular}}
\caption{Training time and prediction accuracy for CRO PUFs}\label{tab8}
   \centering
\begin{tabular}{|c|c|c|}
\hline
\tabincell{c}{{PUF size}\\{(row $\times$ column)}} & Training time & Accuracy \\
\hline
$3 \times 5$ & 5.096 & 100\%  \\
\hline
$3 \times 7$ & 0.739 & 99.9\%  \\
\hline
$3 \times 9$ & 28.013 & 99.9\%  \\
\hline
$4 \times 5$ & 0.411 & 99.9\%  \\
\hline
$4 \times 7$ & 62.136 & 99.9\%  \\
\hline
$5 \times 5$ & 2.027 & 99.9\%  \\
\hline
$6 \times 5$ & 13.607 & 99.9\%  \\
\hline
$7 \times 5$ &50.721 & 99.9\%  \\
\hline
\end{tabular}
\end{table}

\subsection{Key-sharing}

\subsubsection{Extracting of delay matrix}
We extract the delay matrix for CRO PUFs with different sizes using the machine learning. The experiment is conducted on an Intel (R) Core i5-3230M CPU. We have extracted the delay matrices of different CRO PUF sizes such as $3 \times 5, 3 \times 7, 3 \times 9, 4 \times 5, 4 \times 7, 5 \times 5, 6 \times 5$ and $7 \times 5$. In our experiments, all delay paths and the corresponding numbers in the counters are used as the training data, and all CRPs are used as the testing data. TABLE \ref{tab8} gives the training time and the accuracy of the delay matrix extracted from CRO PUFs with different sizes. The accuracy is computed by comparing the matching rate of all CRPs generated on the original delay matrix and predicted delay matrix. Experimental results show that the predicted delay matrix can achieve 99.9\% accuracy. We extracted the delay matrix from a $4 \times 5$ CRO PUF spending only 0.411s with 99.9\% accuracy. For the $3 \times 5$ CRO PUF, we increased the training time to 5.096s to achieve 100\% accuracy.

\subsubsection{Coefficient of stabilization}
In order to verify the usability of CRO PUF-based key-sharing method, we need to know how many reliable CRPs for a PUF with a specified appropriate threshold. In the experiment, we define the coefficient of stabilization (COS) to evaluate the percentage of reliable CRPs.

\begin{equation}
COS = R_{t} / R
\end{equation}
where $R_t$ is the number of CRPs that satisfy the threshold condition, and $R$ denotes the total number of CRPs.

\begin{figure}[!t]
\centering\includegraphics[width=\linewidth]{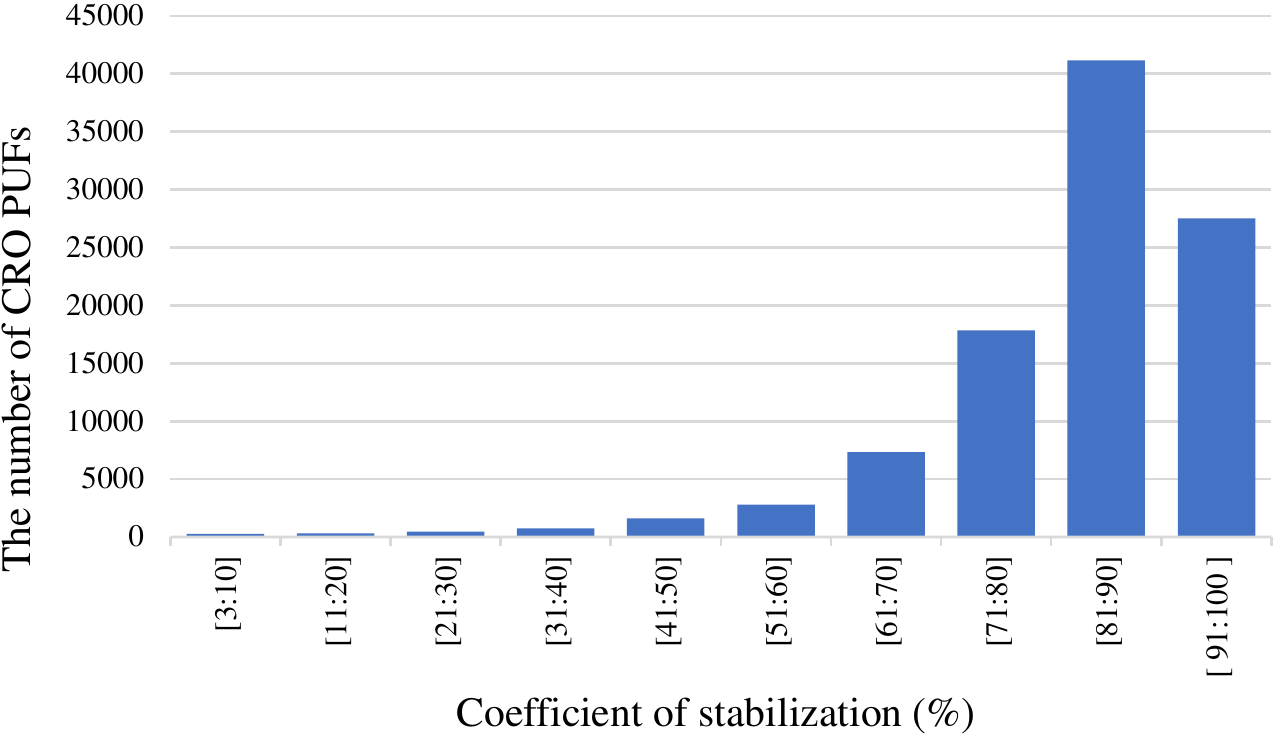}
\caption{COS distribution}\label{fig:13}
\end{figure}

The COS determines the number of reliable CRPs generated by the predicted delay matrix that we extracted. In the experiment, we computed the COSs of $100,000$ simulated $4 \times 5$ CRO PUFs in three FPGA chips. The experimental results are shown in Fig. \ref{fig:13}. The ordinate represents the number of CRO PUFs, and the abscissa represents the COS. For example, [41:50] means that the COS is between 41\% and 50\%. For a $4 \times 5$ CRO PUF, $10,368$ different CRPs can be generated. Assuming that the COS of this CRO PUF is 50\%, there are still $5184$ CRPs that are reliable at different temperatures. As a weak PUF, CRO PUF can generate at least $100$ reliable CRPs even with COS = 1\%. We can see from Fig. \ref{fig:13} that very few COSs are located in the range of $3\% \sim 50\%$ in our experiments, which indicates that available CRPs are enough for key-sharing.

\section{Conclusion}
In many embedded systems and IoT applications, resources-limited devices cannot afford the classic cryptographic security solutions. Lightweight security primitives are required. PUF is an alternative solution for low cost key generation. In this paper, we propose the first PUF-based key-sharing method that the same shared-key can be generated in physically for all devices so that it can be applied in the lightweight key-sharing protocol for IoT devices. CRO PUF structure can effectively improve the reliability and increase hardware efficiency. By selecting different inverters in ROs, the frequency difference between two ROs will be larger than the threshold, and hence generate reliable responses for key sharing. In addition, the lightweight PUF-based key-sharing is not specially designed for CRO PUF but also can be applied to all other configurable PUF structures.


\begin{IEEEbiography}[{\includegraphics[width=1in,height=1.25in,clip,keepaspectratio]{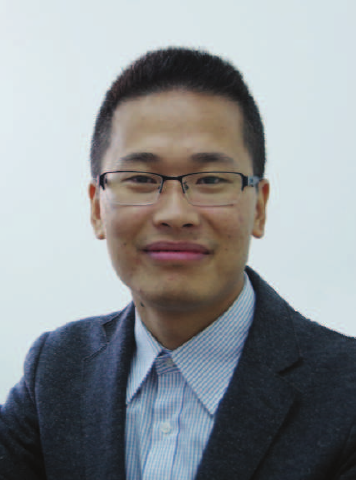}}]{Jiliang Zhang} received the Ph.D. degree in Computer Science and Technology from Hunan University, Changsha, China in 2015. From 2013 to 2014, he worked as a Research Scholar at the Maryland Embedded Systems and Hardware Security Lab, University of Maryland, College Park. From 2015 to 2017, he was an Associate Professor with Northeastern University, China. Since 2017, he has joined Hunan University. His current research interests include hardware/hardware-assisted security, artificial intelligence security, and emerging technologies.

Prof. Zhang is a recipient of the Hu-Xiang Youth Talent, and the best paper nominations in International Symposium on Quality Electronic Design 2017. He has been serving on the technical program committees of many international conferences such as ASP-DAC, FPT, GLSVLSI, ISQED and AsianHOST, and is a senior member of IEEE and a Guest Editor of the Journal of Information Security and Applications and Journal of Low Power Electronics and Applications.
\end{IEEEbiography}

\begin{IEEEbiography}[{\includegraphics[width=1in,height=1.25in,clip,keepaspectratio]{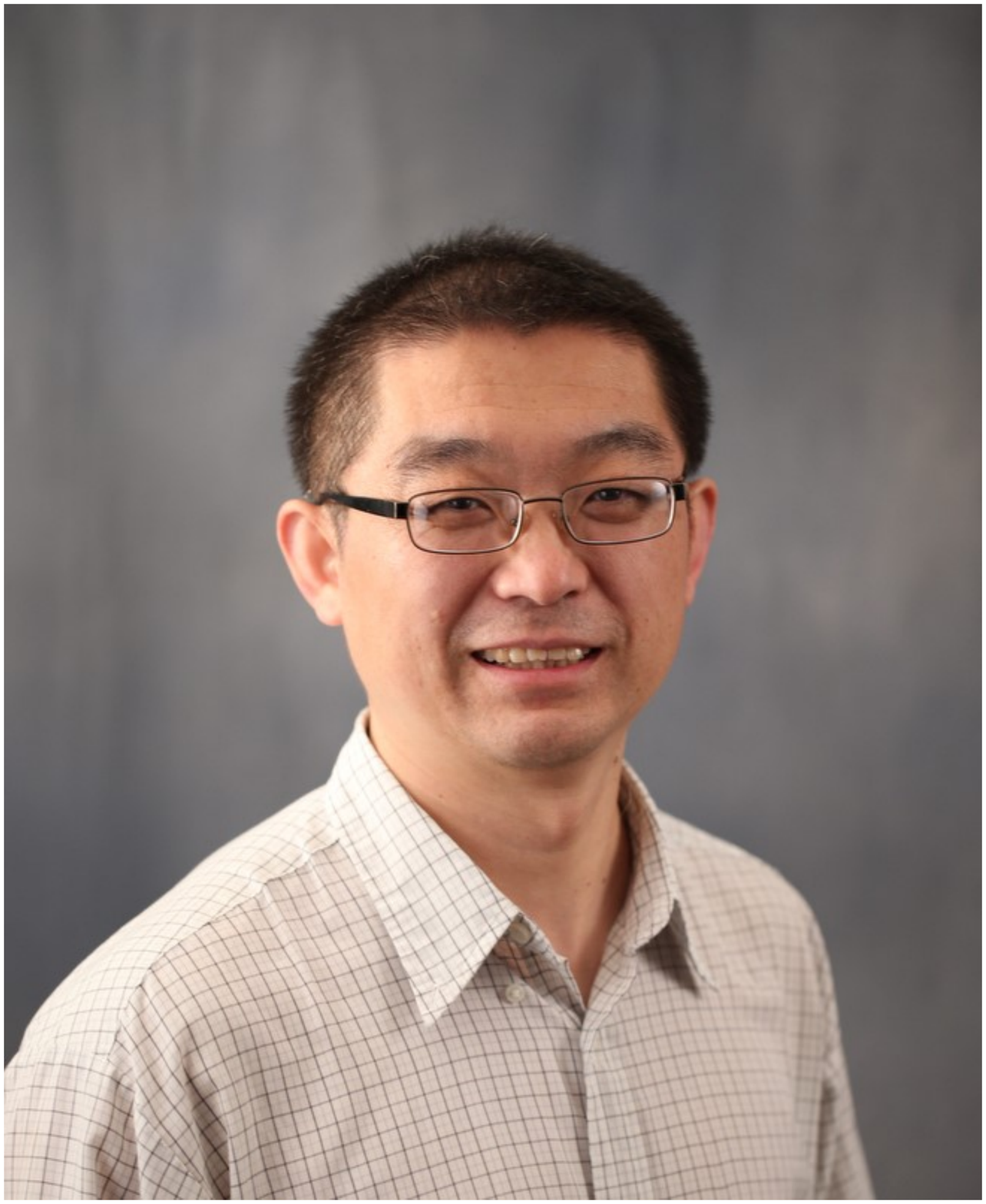}}]{Gang Qu} received the B.S. and M.S. degrees in mathematics from the University of Science and Technology of China, in 1992 and 1994, respectively, and the Ph.D. degree in computer science from the University of California, Los Angeles, in 2000. Upon graduation, he joined the University of Maryland at College Park, where he is currently a professor in the Department of Electrical and Computer Engineering and Institute for Systems Research.

His primary research interests are in the area of embedded systems and VLSI (Very Large Scale Integration) CAD (Computer Aided Design) with focus on low power system design and hardware related security and trust. He studies optimization and combinatorial problems and applies his theoretical discovery to applications in VLSI CAD, wireless sensor network, bioinformatics, and cybersecurity. Dr. Qu has received many awards for his academic achievements, teaching, and service to the research community. He is a senior member of IEEE and serving as associate editor for the IEEE Transactions on Computers, IEEE Embedded Systems Letters and Integration, the VLSI Journal.
\end{IEEEbiography}

\vfill

\end{document}